\documentclass{ifacconf}

\usepackage{graphicx}      
\usepackage{natbib}        

\usepackage{amssymb,color,amsmath,graphicx}

\newcommand{\cI} { {\mathcal I}}
\newcommand{\cJ} { {\mathcal J}}
\newcommand{\cK} { {\mathcal K}}
\newcommand{\cX} { {\mathcal X}}
\newcommand{\cY} { {\mathcal Y}}
\newcommand{\cU} { {\mathcal U}}

\newcommand{\bR} { {\mathbb R}}
\newcommand{\intr}[1] {{\rm int}(#1)}
\def\red{\hfill $\lhd$}

\begin{document}
\begin{frontmatter}

\title{Scalable Control Design for \\ $\cK$-positive Linear Systems\thanksref{footnoteinfo}} 

\thanks[footnoteinfo]{This work of Kawano was supported in part by JSPS KAKENHI Grant Number JP19K23517.}

\author[First]{Yu Kawano} 
\author[Second]{Fulvio Forni}

\address[First]{Graduate School of Engineering, Hiroshima University, 
Kagamiyama 1-4-1, Higashi-Hiroshima 739-8527, Japan (e-mail: ykawano@hiroshima-u.ac.jp).}
\address[Second]{Department of Engineering, University of Cambridge,
Cambridge, CB2 1PZ, UK (e-mail: f.forni@eng.cam.ac.uk)}

\begin{abstract}                
Systems whose variable are constrained to be positive allow computationally efficient control design. 
We generalize these results to linear systems which leave a cone invariant. This is a wider class of systems than positive systems. We revisit classical results on stability and dissipativity of positive linear systems and show how scalable conditions on linear programming can be extended to cone invariant linear systems. Our results are illustrated by scalable stabilizing controller design for mass-spring systems.
\end{abstract}

\begin{keyword}
positivity, controller design, scalability, linear programming
\end{keyword}

\end{frontmatter}
\section{Introduction}
Linear systems for which the positive orthant is a forward invariant set are called positive systems. 
Positivity naturally arises in biological networks, social networks, and in all those systems whose state variables 
typically represent non-negative physical quantities (concentrations, populations, magnitudes). 
Positive linear systems are monotone systems with respect to the component-wise order (\cite{Smith:08,Hirsch2003,AS:03})
and, notably, Perron-Frobenius theory shows that these systems have a dominant mode
constraining their asymptotic behavior to a one-dimensional ray (\cite{Farina2000}).
This property strongly simplifies stability and dissipativity analysis, which reduce 
to finding linear Lyapunov or storage functions instead of quadratic ones; see e.g.~\cite{Rantzer:15,HCH:10}. 
Motivated by these facts, scalable analysis and control methods have been intensively studied; see e.g.~\cite{Rantzer:15TAC,Rantzer:15,EPA:16,TL:11,RT:07}. The aim of this paper is to show that scalable control methods
extend to any linear system that admits a forward invariant (solid, convex, pointed) cone $\cK$, 
not necessarily corresponding to the positive orthant. We call these systems $\cK$-positive, to avoid ambiguities.

Taking advantage of the well-established theory of $\cK$-positive systems (\cite{BP:94,Bushell1973}),
we show how $\cK$-positivity can replace classical positivity in systems analysis, 
opening the way to scalable analysis and simplified control methods 
to a much wider class of systems. For example, a network of mass-spring systems is 
not positive in the classical sense but can be shown to be $\cK$-positive for a wide range 
of parameters, therefore amenable to scalable stability/dissipativity 
analysis based on linear programming (LP). 

Taking inspiration from the fact that the positive orthant is a specific polyhedral proper cone~(\cite{BP:94}),
we focus on $\cK$-positive linear systems with respect to polyhedral proper cones $\cK$. We show that the aforementioned scalable stability and dissipativity analysis for positive linear systems can be equivalently delivered to $\cK$-positive linear systems by employing linear Lyapunov and storage functions. Furthermore, 
we derive stability conditions for the interconnection of dissipative $\cK$-positive linear systems. These results can be 
also used for scalable controller design, as illustrated by an example of mass-spring controlled systems.

For reason of space our analysis is limited to linear systems but the theory can be extended to monotone systems via differential analysis / linearization,
following the approach of \cite{FS:16}. We leave this to future publications.
The rest of the paper is organized as follows. Section~\ref{sec:FPCI} provides a motivating example for investigating $\cK$-positive systems. Section~\ref{sec:positive} summarizes existing scalable results on positive systems. 
Section~\ref{sec:con_inv} generalizes the results of Section~\ref{sec:positive} to $\cK$-positive systems.  Examples are provided in Section~\ref{sec:ex}. Conclusions follows. Proofs are in appendix.

{
\textbf{Notation:} Let~$\bR^n$ and~$\bR_+^n$ be the sets of real numbers and non-negative real numbers, respectively. A closed subset~$\cK \subset \bR^n$ is called a \emph{proper cone} if this satisfies the following four properties: 1) a cone, i.e.,~$a \cK \subset \cK$ for any~$a \in \bR_+$, 2) convex, i.e.,~$\cK + \cK \subset \cK$, 3) pointed, i.e.~$-\cK\cap \cK =\{0\}$, and 4) solid, i.e.,~$\intr \cK \neq \emptyset$. A proper cone~$\cK$ induces a partial order in~$\bR^n$ via~$x \preceq x'$ if and only if~$x' - x \in \cK$. In addition we use the notation~$x \prec x'$ if and only if $x' - x \in \intr{\cK}$.

A cone~$\cK$ is said to be \emph{polyhedral} if there exists a finite set of vectors (or called generators)~$K_i \in \bR^n \setminus \{0\}$,~$i \in \cI_{\cK} :=\{1,\dots,p_{\cK}\}$ such that
\begin{align*}
\cK = \left\{ x \in \bR^n: x = \sum_{i \in \cI_{\cK}} a_i K_i, \; a_i \ge 0\right\}.
\end{align*}
The \emph{dual cone} of a proper cone~$\cK \subset \bR^n$ is defined by
\begin{align*}
\cK^*:=\{z \in \bR^n: x \in \cK \implies x^\top z \ge 0\},
\end{align*}
where~$x^\top$ denotes the transpose of~$x$. Its interior is
\begin{align*}
\intr{\cK^*}=\{z \in \bR^n: x \in \cK \setminus \{0\} \implies x^\top z > 0\}.
\end{align*}
If~$\cK$ is a convex cone,~$\cK=\cK^{**}$. For a proper or polyhedral cone~$\cK$, its dual cone is also proper or polyhedral, respectively. Therefore, 
it is generated by a finite set of vectors~$K^*_i \in \bR^n \setminus \{0\}$,~$i \in \cI_{\cK^*}: =\{1,\dots,p_{\cK^*}\}$. This leads to 
the equivalent representation of~$\cK$ as
\begin{align*}
\cK = \{x \in \bR^n:  K^* x \ge 0\}, \;
(K^*)^\top := \begin{bmatrix} K^*_1 & \cdots &K^*_r \end{bmatrix}.
\end{align*}

In this paper, a triplet~$(\cX,\cU,\cY)$ denotes polyhedral proper cones corresponding to the state, input, and output. Their dual cones are denoted by~$(\cX^*,\cU^*,\cY^*)$. Moreover, generators of each cone are denoted by using the corresponding roman font. For instance, generators of~$\cX$ are denoted by~$X_{i_\cX}$,~$i_\cX \in \cI_\cX:= \{1,\dots, p_{\cX}\}$. For the sake of simplicity, $\forall i_\cX \in \cI_\cX$ is described as~$\forall i_\cX$.
}

\section{From Positivity to $\cK$-positivity: motivating example}\label{sec:FPCI}
Consider the linear mass-spring system described by
\begin{align}
\begin{bmatrix}
\dot x_1 \\ \dot x_2
\end{bmatrix}
&=
\begin{bmatrix}
0 & 1 \\
-k/m & 0
\end{bmatrix}
\begin{bmatrix}
x_1 \\ x_2
\end{bmatrix}
+
\begin{bmatrix}
0 \\ 1
\end{bmatrix} u, \quad m,k>0,
\nonumber\\
y &= x_1,
\label{sys:mech}
\end{align}
where~$x_1$,~$x_2 \in \bR$ denote the position and velocity of the point mass, respectively, and~$u$,~$y\in\bR$ denote the exogenous input and sensor measurement, respectively. 

We design of a feedback controller $u=F_1 x_1 + F_2  x_2$
which makes the closed-loop system positive and exponentially stable. Positivity requires
\begin{align}
F_1 -k/m \ge 0 
\label{cond:mech_p}
\end{align} 
since the off-diagonal elements of the state matrix of a positive system must be non-negative. 
For asymptotic stability, the eigenvalues of the closed-loop 
\begin{align*}
\lambda = \frac{F_2 \pm \sqrt{F_2^2 + 4 (F_1 - k/m)}}{2}
\end{align*} 
have negative real part if and only if
\begin{align}
\left\{\begin{array}{l}
F_2<0,\\
F_1 - k/m<0,
\end{array}\right.\label{cond:mech_s}
\end{align}
which show that positivity and asymptotic stability are incompatible for the controlled mass-spring system.

For this specific example, positivity enables scalable control design (\cite{Rantzer:15}) but 
also makes asymptotic stability unfeasible. However, scalability can be retained by using a 
different cone, $\cK$, generated by 
\begin{align}
K_1 :=  \begin{bmatrix} 0 \\ 1 \end{bmatrix}, \quad
K_2:= \begin{bmatrix} 1 \\ -2 \end{bmatrix} ,
\label{eq:mech_cone}
\end{align}
and wider than the positive orthant.

According to Theorem~\ref{thm:cone_inv} below, $\cK$ is a forward invariant set of the closed-loop system if and only if
\begin{align*}
0 \leq &\begin{bmatrix} 1 & 0 \end{bmatrix}
\begin{bmatrix}
0 & 1 \\
F_1 - k/m & F_2
\end{bmatrix}
\begin{bmatrix} 0 \\ 1 \end{bmatrix}
= 1 , \\
0 \leq &
\begin{bmatrix} 2 & 1 \end{bmatrix}
\begin{bmatrix}
0 & 1 \\
F_1 - k/m & F_2
\end{bmatrix}
\begin{bmatrix} 1 \\ -2 \end{bmatrix}
= - 4 - k/m + F_1 - 2 F_2  \ .
\end{align*}
Notably, the inequality above is linear in $F_1$ and $F_2$ (like \eqref{cond:mech_p}).
Furthermore we can rewrite the inequality above as
\begin{align}
F_2 \le (- 4 + F_1 - k/m )/2
\label{cond:mech_kpos}
\end{align} 
which shows that 
\eqref{cond:mech_kpos} and \eqref{cond:mech_s} is a feasible set of \emph{linear constraints}
in $F_1$ and $F_2$,
ensuring $\cK$-positivity and asymptotic stability of the controlled mass-spring system. 
Motivated by this example, in what follows we revisit and extend the scalable approaches in \cite{Rantzer:15,HCH:10})
to $\cK$-positive systems.

\section{Revisit Scalable Stability/Dissipativity Analysis for Positive Systems}\label{sec:positive}
We briefly revisit here the main results in~\cite{Rantzer:15,HCH:10} for stability and dissipativity of positive linear systems. 
Consider the linear system,
\begin{align}
\Sigma: \left\{\begin{array}{l}
\dot x = A x + B u,\\
y = C x,
\end{array}\right.
\label{eq:sys_gen}
\end{align}
where~$x \in \bR^n$,~$u \in \bR^m$, and~$y \in \bR^p$ denote the state, input, and output, respectively, and~$A \in \bR^{n \times n}$,~$B \in \bR^{n \times m}$, and~$C \in \bR^{n \times p}$. Suppose that the system is positive, that is, for any given 
state trajectory and input-output signals
$(x(\cdot ),u(\cdot),y(\cdot))$, if $x(0)  \!\in \! \bR_+^n$ and $u(t)  \!\in \! \bR_+^m$ for all $t \!\geq \! 0$ then 
$(x(t),y(t))   \!\in \! \bR_+^n  \!\times \! \bR_+^p$ for all $t \!\geq \! 0$. 
A necessary and sufficient condition for positivity is that every off-diagonal element 
of~$A$ and every element of~$B$ and~$C$ are non-negative.

For positive linear systems, the following scalable stability condition is well known; (see e.g.,~)\cite{Rantzer:15,HCH:10}.
\begin{prop}~\cite[Theorem 2.11.]{HCH:10}\label{prop:p_s}
A positive LTI system is Hurwitz if and only if there exists~$v \in \bR^n$ such that~$v \succ 0$ and~$- v^\top A \succ 0$. \red
\end{prop}

Proposition~\ref{prop:p_s} shows that stability for positive systems is an LP problem, whose complexity scales linearly with the 
size of the system. Stability (and scalability) follows from the fact that the linear function $V(x)=v^\top x$ is
a Lyapunov function for the system whenever $x\in \bR^n_+$.

Moving from quadratic to linear storage functions and supply rates, a positive system is exponentially dissipative with respect
to the supply rate $s(u,y) = r^\top u + q^\top y$, $r \in \bR^m$, $q \in \bR^p$, if there exist
$\alpha>0$ and $v \in \bR_+^n$ such that 
\begin{align*}
e^{\alpha t_2} v^\top x(t_2) \le e^{\alpha t_1} v^\top x(t_1) + \int_{t_1}^{t_2} e^{\alpha t} (r^\top u(t) + q^\top y(t)) dt, 
\end{align*}
for any~$[t_1,t_2]\subseteq \bR_+$ and any $(x(\cdot),u(\cdot),y(\cdot))$ such that $(x(t),u(t),y(t)) \in \bR_+^n \times \bR_+^m \times \bR_+^p$ for all $t \in [t_1,t_2]$.

Exponential dissipativity of positive LTI systems can be verified as follows. 
\begin{prop}~\cite[Theorem 5.3.]{HCH:10}\label{prop:p_d}
A positive LTI system~$\Sigma$ is exponentially dissipative with respect to supply rate~$r^\top u + q^\top y$ with a continuously differentiable storage function if and only if there exist~$\alpha \!> \!0$ and~$v  \!\succ \! 0$ such that
$q^\top  \! C \succeq v^\top  \! A + \alpha v^\top$ and $r^\top \! \succeq v^\top  \! B$. \red
\end{prop}

Finally, using $\Sigma$ and the new system
\begin{align}
\Sigma_c: \left\{\begin{array}{l}
\dot x_c = A_c x_c + B_c u_c,\\
y_c = C_c x_c.
\end{array}\right.
\label{eq:contr_gen}
\end{align}
the following proposition clarifies the use of (linear) exponential dissipativity for closed-loop interconnections.
\begin{prop}~\cite[Theorem 5.5.]{HCH:10}\label{prop:p_ic_s}
Suppose that the two positive systems~$\Sigma$ and~$\Sigma_c$ are 
exponentially dissipative with respect to supply rates~$r^\top u + q^\top y$ and~$r_c^\top u_c + q_c^\top y_c$, 
respectively. Their positive feedback interconnection $u_c=y$ and~$u=y_c$ is exponentially stable 
if there exist scalars $\sigma >0$ and $\sigma_c > 0$ such that 
$-q \succeq \sigma_c r_c$ and $-  r \succeq \sigma q_c$. \red
\end{prop}

Notably the number of constraints in Propositions \ref{prop:p_d} and \ref{prop:p_ic_s} 
grows linearly with the size of the system, opening the way to scalable analysis
of large large network systems. The generalization to $\cK$-positive systems in the next
section will retains both the important features of constraint linearity (LP) and scalability.

\section{Scalable Stability/Dissipativity Analysis for $\cK$-positive Systems}\label{sec:con_inv}

\subsection{$\cK$-positivity and stability}
\begin{defn}
The linear system~$\Sigma$ is said to be positive
with respect to polyhedral proper cones~$(\cX,\cU,\cY)$ if,
for any given 
state trajectory and input-output signals $(x(\cdot ),u(\cdot),y(\cdot))$,
\begin{itemize}
\item[] if $x(0)  \in \cX$ and $u(t)  \in  \cU$ for all $t  \geq  0$, \vspace{1mm}
\item[] \hspace{1.5cm}then $(x(t) ,y(t))  \in \cX \times \cY$ for all $t \!\geq \! 0$.   \red
\end{itemize}
\end{defn}
For closed systems positivity with respect to the cone $\cX$ simply means that,
along any trajectory $x(\cdot)$,
$x(0) \in \cX$ implies $x(t) \in \cX$ for all $t\geq 0$.

The next theorem shows that $\cK$-positivity is tractable. 
\begin{thm}\label{thm:cone_inv}
The linear system~$\Sigma$ is positive
with respect to polyhedral proper cones~$(\cX,\cU,\cY)$ 
if and only if 
\begin{enumerate}
\renewcommand{\theenumi}{\Roman{enumi}}
\item $e^{At} \cX \subseteq \cX$ for $t\geq 0$, i.e.
\begin{align}
(X^*_{i_{\cX^*}})^\top A X_{i_\cX} \ge 0 \label{cond1:cone_inv}
\end{align}
for all~$i_{\cX^*}$ and~$i_{\cX}$ 
such that~$(X^*_{i_{\cX^*}})^\top X_{i_\cX} = 0$; 
\item $B \cU \subseteq \cX$, i.e.
\begin{align}
(X^*_{i_{\cX^*}})^\top B U_{i_\cU} \ge 0
\label{cond2:cone_inv}
\end{align}
for all~$i_{\cX^*}$ and~$i_\cU$;
\item $C \cX \subseteq \cY$, i.e.
\begin{align}
(Y^*_{i_{\cY^*}})^\top C X_{i_\cX} \ge 0
\label{cond3:cone_inv}
\end{align}
for all~$i_{\cX}$ and~$i_{\cY^*}$.
\red
\end{enumerate}
\end{thm}

$\cK$-positive systems allow for simplified Lyapunov analysis.
\begin{thm}\label{thm:cone_s}
Suppose that $\Sigma$ is positive with respect to  polyhedral proper cones~$(\cX,\cU,\cY)$.
$\Sigma$ is exponentially stable if and only if there exists~$v \in \bR^n$ such that
\begin{subequations}\label{cond:cone_s}
\begin{align}
v^\top X_{i_\cX} > 0&,
\label{cond1:cone_s}\\
- v^\top A X_{i_\cX} > 0&
\label{cond2:cone_s}
\end{align}
\end{subequations}
for all~$i_{\cX}$. \red
\end{thm}
\eqref{cond1:cone_s} guarantees $v \in \intr{\cX^*}$ therefore $v^\top x$ is a Lyapunov function for all $x \in \cX$,
and \eqref{cond2:cone_s} guarantees the exponential decay of the Lyapunov function. 
Together, \eqref{cond1:cone_inv} and~\eqref{cond:cone_s} certify positivity with respect to $\cX$ and asymptotic stability of the autonomous system $\dot{x} = Ax$ via linear programming.

\subsection{Dissipativity}
In this subsection, we generalize Propositions~\ref{prop:p_d} and~\ref{prop:p_ic_s}
to $\cK$-positive systems.  
 \begin{defn}
Let $\Sigma$ be positive with respect to polyhedral proper cones~$(\cX,\cU,\cY)$.
$\Sigma$ is  \emph{exponentially dissipative} with respect to supply rate~$s:\cU \times \cY \to \bR$
given by $s(u,y) = r^\top u + q^\top y$, $r \in \bR^m$, $q \in \bR^p$, if there exist
$\alpha>0$ and $v \in \cX^*$ such that 
\begin{align*}
e^{\alpha t_2} v^\top x(t_2) \le e^{\alpha t_1} v^\top x(t_1) + \int_{t_1}^{t_2} e^{\alpha t} (r^\top u(t) + q^\top y(t)) dt, 
\end{align*}
for any~$[t_1,t_2]\subseteq \bR_+$ and any state trajectory and input-output signals 
$(x(\cdot),u(\cdot),y(\cdot))$ such that $(x(t),u(t),y(t)) \in \cX\times\cU\times\cY$ for all $t \in [t_1,t_2]$. \red

\end{defn}
The above dissipation inequality is equivalent to
\begin{align}
\frac{d  (v^{\!\top} \! x(t))}{dt} + \alpha v^\top x(t) \le  r^\top u(t) + q^\top y(t), \; \forall t \in \bR_+
\label{defn:ed}
\end{align}
for all~$(x(t ),u(t ),y(t )) \in \cX \times \cU \times \cY$, which leads
to the following conditions (generalization of Proposition~\ref{prop:p_d}).
\begin{thm}\label{thm:cone_d}
Let $\Sigma$ be positive with respect to polyhedral proper cones~$(\cX,\cU,\cY)$.
$\Sigma$ is exponentially dissipative with respect to supply rate~$s(u,y) = q^\top y + r^\top u$
if and only if there exist~$\alpha>0$ and~$v\in \bR^n$ such that
\begin{subequations}\label{cond:cone_d}
\begin{align}
v^\top X_{i_\cX} &\ge 0, \label{cond1:cone_d}\\
q^\top C X_{i_\cX}& \ge ( v^\top A + \alpha v^\top ) X_{i_\cX}, \label{cond2:cone_d}\\
r^\top U_{i_\cU} &\ge v^\top B  U_{i_\cU} \label{cond3:cone_d}
\end{align}
\end{subequations}
for all~$i_{\cX}$ and~$i_\cU$.
\red
\end{thm}

The restriction to positive storages $v^\top x>0$ for $0 \neq x \in \cX$ is 
required if one wants to connect exponential dissipativity to closed-loop exponential stability
without additional observability assumptions. In practice, \eqref{cond1:cone_s} replaces \eqref{cond1:cone_d}.
Also, using \eqref{cond1:cone_s} allows to replace \eqref{cond2:cone_d} with the simpler condition\footnote{
\eqref{cond22:cone_d} follows from~$\alpha > 0$,~\eqref{cond1:cone_s} and~\eqref{cond2:cone_d}.
Conversely, if there exists~$v \in \bR^n$ satisfying~\eqref{cond1:cone_s} and~\eqref{cond22:cone_d}, then there exists~$\alpha_{i_\cX}>0$ such that
$
q^\top C X_{i_\cX} \ge ( v^\top A + \alpha_{i_\cX} v^\top ) X_{i_\cX}.
$
Therefore,~\eqref{cond2:cone_d} holds for~$\alpha := \min_{i_{\cX}}\{\alpha_{i_\cX}\}$. 
} 
\begin{align}
q^\top C X_{i_\cX} > v^\top A X_{i_\cX}  
\label{cond22:cone_d} \ .
\end{align}
Notably,~\eqref{cond1:cone_s},~\eqref{cond3:cone_d}, and~\eqref{cond22:cone_d} show that exponential
dissipativity for $\cK$-positive systems is an LP problem in the variables ~$v$,~$q$, and~$r$.

\subsection{Stability of Networked Interconnected Systems}
We study the exponential stability of closed systems
arising from the network interconnection of positive systems $\Sigma_j$ with respect 
to polyhedral proper cones~$(\cX_j,\cU_j,\cY_j)$, $j \in \cJ_N:=\{1,\dots,N\}$. Each system is 
represented by 
\begin{align*}
\Sigma_j: 
\left\{\begin{array}{l}
\dot x_j = A_j x_j + B_j u_j,\\
y_j = C_j x_j,
\end{array}\right.
\end{align*}
where~$x_j \in \bR^{n_j}$,~$u_j \in \bR^{m_j}$, and~$y_j \in \bR^{p_j}$,~$j \in \cJ_N$. 
Interconnections are represented by 
\begin{align}
u_j = \sum_{k \in \cJ_N} W_{j,k} y_k, \quad j \in \cJ_N.
\label{eq:fbic}
\end{align}
The following result generalizes Proposition~\ref{prop:p_ic_s} to closed networks of 
$\cK$-positive systems. 
\begin{thm}\label{thm:cone_ic_s}
Let $\Sigma_j$ be positive with respect to polyhedral proper cones $(\cX_j,\cU_j,\cY_j)$, $j \in \cJ_N$.
Suppose that
\begin{enumerate}
\renewcommand{\theenumi}{\Roman{enumi}}
\item $W_{j,k} \cY_k \subset \cU_j$, i.e.
\begin{align}
(U^*_{j,i_{\cU_j^*}})^\top W_{j,k} Y_{k,i_{\cY_k}} \ge 0,
\label{cond1:cone_ic_s}
\end{align}
for all~$i_{\cU_j^*}$,~$i_{\cY_k}$ and~$j,k\in \cJ_N$;
\item each subsystem is exponentially dissipative with respect to supply rates~$r_j^\top u_j+ q_j^\top y_j$ with positive definite storage function~$v_j^\top x_j$ in~$\cX_j$; 
\item there exist~$\sigma_j>0$,~$j \in \cJ_N$, such that
\begin{align*}
- \sum_{j \in \cJ_N} \sigma_j  q_j^\top Y_{j,i_{\cY_j}}  \ge   \sum_{j \in \cJ_N} \sigma_j r_j^\top  \sum_{k \in \cJ_N} W_{j,k} Y_{k,i_{\cY_k}} \ 
\end{align*}
for any combination of generators 
$Y_{j,i_{\cY_j}}$ and $Y_{j,i_{\cY_k}}$.
\end{enumerate}
Then, the interconnected system arising from \eqref{eq:fbic} is positive with respect to~$\cX = \cX_1 \times \cdots \times \cX_N$ and 
exponentially stable. \red
\end{thm}

Theorem~\ref{thm:cone_ic_s} leads to a LP formulation for the stability of the interconnected system.
Define~$\bar v_j := \sigma_j v_j$,~$\bar q_j := \sigma_j q_j$, and~$\bar r_j := \sigma r_j$. Using 
\eqref{cond1:cone_s},~\eqref{cond3:cone_d}, and~\eqref{cond22:cone_d} for each subsystem we have that (II)
is satisfied if 
\begin{subequations}\label{cond2:cone_ic_s}
\begin{align}
\bar v_j^\top X_{j,i_{\cX_j}} &> 0, \\
\bar q_j^\top C_j X_{j,i_{\cX_j}}  &> \bar v_j^\top A_j X_{j,i_{\cX_j}},\\
\bar r_j^\top U_{j,i_{\cU_j}} &\ge \bar v_j^\top B_j U_{j,i_{\cU_j}}
\end{align}
\end{subequations}
for all~$i_{\cX_j}$,~$i_{\cU_j}$, and~$j \in \cJ_N$. 
Moreover,  (III) is satisfied if
\begin{align}
- \sum_{j \in \cJ_N} \bar  q_j^\top Y_{j,i_{\cY_j}}  \ge   \sum_{j \in \cJ_N} \bar r_j^\top \sum_{k \in \cJ_N}  W_{j,k} Y_{k,i_{\cY_k}}.
\label{cond3:cone_ic_s}
\end{align}
for any combination of generators 
$Y_{j,i_{\cY_j}}$ and $Y_{j,i_{\cY_k}}$.

\section{Examples for Mass-Spring Systems}
\label{sec:ex}
\subsection{$\cK$-positivity and stabilization of a single system}
\label{sec:example_single}
We revisit the mechanical system in Section~\ref{sec:FPCI}. We design 
a state feedback controller~$u=Fx$ imposing both $\cK$-positivity and exponential stability. For simplicity, we consider mass $m=1$ 
and spring constant $k=1$. 

Consider the cone~$\cX$ generated by~\eqref{eq:mech_cone}. Its dual cone~$\cX^*$ is generated by
\begin{align*}
X^*_1:=\begin{bmatrix} 1 \\ 0 \end{bmatrix}, \quad
X^*_2:=\begin{bmatrix} 2 \\ 1 \end{bmatrix}.
\end{align*}

The condition for forward invariance~\eqref{cond1:cone_inv} reads
\begin{align}
(X^*_{i_{\cX^*}})^\top (A + B F) X_{i_\cX} \ge 0, \quad i_\cX=i_{\cX^*}
\label{cond:ex_cone}
\end{align}
Next, the stability conditions~\eqref{cond:cone_s} become
\begin{subequations}
\label{cond2:ex_cone}
\begin{align}
v^\top X_{i_\cX} > 0,&  \quad i_\cX =1,2, \label{cond2:ex_conea}\\
- v^\top (A + B F) X_{i_\cX} > 0,& \quad i_\cX =1,2 \label{cond2:ex_coneb}.
\end{align}
\end{subequations}
For any fixed $v\in \cX$, conditions \eqref{cond:ex_cone} and \eqref{cond2:ex_cone}
define a linear program in $F$, whose feasibility leads to a stabilizing controller that
enforces positivity with respect to $\cX$. 

Although \eqref{cond:ex_cone} and \eqref{cond2:ex_cone} is not a linear program 
with respect to $F$ and~$v$, these conditions can be further adapted into a LP formulation.
For instance, any~$v \in $ satisfying \eqref{cond2:ex_conea}  can be described as
\begin{align}
v = a_1 X^*_1
+ a_2 X^*_2, \; a_1,a_2 >0,
\label{eq:ex_v}
\end{align}
which combined to \eqref{cond:ex_cone} leads to
\begin{align}
- (a_1 X^*_1+ a_2 X^*_2)^\top (A + B F) X_{i_\cX} > 0, \quad i_\cX =1,2.
\label{cond1:ex_s}
\end{align}
This is still not a set of LP problems with respect to~$a_1,a_2>0$ and~$F$. 
However, feasibility of \eqref{cond:ex_cone} and~\eqref{cond1:ex_s} requires
\begin{align}
- (X^*_{i_{\cX^*}})^\top (A + B F) X_{i_\cX} > 0, \quad i_\cX  \neq i_{\cX^*} .
\label{cond2:ex_s}
\end{align}
Indeed, \eqref{cond:ex_cone} and~\eqref{cond2:ex_s} are LP problems with respect to~$F$. Therefore, first we solve this LP problem. Then,~\eqref{cond1:ex_s} becomes LP problems with respect to~$a_1,a_2>0$. 

\begin{figure}
\begin{center}
\includegraphics[width=70mm]{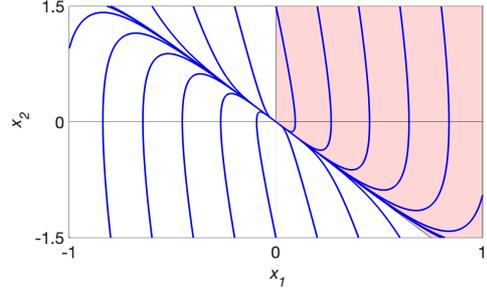} 
\caption{Phase portrait of the closed-loop system. The red region represents the cone $\cX$.} 
\label{fig:mech_stab}
\end{center}
\end{figure}

By solving the above separated LP problems, we obtain~$F = \begin{bmatrix} -21.0 & -13.0 \end{bmatrix}$ and~$v = \begin{bmatrix} 5.09 & 0.836 \end{bmatrix}^\top$. Figure~\ref{fig:mech_stab} shows the phase portrait of the closed-loop system. The state feedback $u=Fx$ guarantees positivity with respect to $\cX$ and exponential stability simultaneously. We emphasize that  the closed-loop system is not positive with respect to $\bR_+^2$.

\subsection{Stabilization of ring network}
We design a stabilizing feedback controller for a ring network topology of (uniformly controlled) mass-spring systems. 
The control input to each subsystem reads
\begin{align*}
u_j = F  x_j + y_{j+1}, \quad j \in \cJ_N,
\end{align*}
where~$y_{N+1} := y_1$. The objective here is to design the uniform feedback gain~$F$ 
such that the interconnected (closed) system is $\cK$-positive and exponentially stable.

As the triplet of cones, we choose~$(\cX,\bR_+,\bR_+)$ for each subsystem, where~$\cX$ is the polyhedral proper cone used in the previous subsection. Then, the cone invariance condition consists of~\eqref{cond:ex_cone} and
\begin{align}
(X^*_{i_{\cX^*}})^\top B &\ge 0, \quad i_{\cX^*} =1,2, \label{condu:ex_cone}\\
C X_{i_\cX} &\ge 0, \quad i_\cX =1,2.\label{condy:ex_cone}
\end{align}
We observer that \eqref{condu:ex_cone} and~\eqref{condy:ex_cone} are trivially satisfied, since $C \in \cX^*$ and $B\in \cX$. 
Item~(I) of Theorem~\ref{thm:cone_ic_s} also holds. Therefore, the feedback gain~$F$ is constrained only by the LP condition \eqref{cond:ex_cone}.

Next, let us consider Item~(II) of Theorem~\ref{thm:cone_ic_s}, namely \eqref{cond2:cone_ic_s} 
We fix every~$\bar v_j$,~$j \in \cJ_N$ to $\bar v_j=v$ where $v$ corresponds to the one found in Section \ref{sec:example_single}.
Since $\bar q_j$ and~$\bar r_j$ are arbitrary, we take the uniform selection $\bar q_j = q$ and~$\bar r_j =r$, $j \in \cJ_N$.
\eqref{cond2:cone_ic_s} reduce to
\begin{subequations}
\label{eq:ex2_cond2}
\begin{align}
q C X_{i_\cX} &> v^\top (A + B F)  X_{i_\cX},\\
r & \ge v^\top B \ .
\end{align}
\end{subequations}

Finally, we consider Item~(III) of Theorem~\ref{thm:cone_ic_s}, namely~\eqref{cond3:cone_ic_s}. From the interconnection structure,~\eqref{cond3:cone_ic_s} becomes 
\begin{align}
\label{eq:ex2_cond3}
- \sum_{j \in \cJ_N} q \ge \sum_{j \in \cJ_N} r,
\end{align}
\eqref{cond:ex_cone}, \eqref{eq:ex2_cond2}, and \eqref{eq:ex2_cond3} define
a linear program in $q$, $r$, and $F$, whose solution leads to a stabilizing state-feedback 
controller $u=Fx$, which also enforces  positivity with respect to $\cX^N$ 
of the closed-loop system.

For instance, taking $r = v^\top B$ and $q= - r$, 
conditions \eqref{cond:ex_cone}, \eqref{eq:ex2_cond2}, and \eqref{eq:ex2_cond3} reduce to 
\begin{align}
-v^\top B C X_{i_\cX} > v^\top (A + B F)  X_{i_\cX}.
\label{cond2:ex_cone_s}
\end{align}
$F= \begin{bmatrix} -21.0 & -13.0 \end{bmatrix}$ solves 
\eqref{cond:ex_cone} and~\eqref{cond2:ex_cone_s} and is a uniform feedback
which stabilizes the ring network for any number $N$ of interconnected systems 
(scale free solution).  For example, for $N=1000$, the maximum real part of the 
eigenvalues of the interconnected closed-loop system is~$-1.48$.

\section{Conclusion}\label{sec:con} 
We have generalized scalable stability and dissipativity conditions to $\cK$-positive systems using linear Lyapunov and storage functions.
Our conditions are linear in the variables of interest, enabling the use of linear programming for stability analysis of $K$-positive systems.
Examples of stabilization and network stabilization of mass-spring systems illustrate how to use our results for scalable control design. 

The main limitation of the proposed approach is in the use of pre-defined cones $\cX$, which play in the current problem the same pivotal role
of control Lyapunov functions in classical stabilization. 
Like for Lyapunov theory, finding a suitable cone is not a trivial matter. Indeed, our motivating example shows how starting with the 
``wrong'' cone (positive orthant) could lead to severe limitations in control design. 
This calls for further investigations on the co-design of feedback control and cone. First attempts in this directions can be found in \cite{Kousoulidis2020}.

The results of the paper show promise in analysis and control of nonlinear systems, by leveraging 
differential positivity \citep{FS:16}  and monotonicity (\cite{Smith:08,AS:03}).  
A related stability analysis method has been developed in~\cite{KBC:20}. 
Using the system linearization, Theorems \ref{thm:cone_inv}, \ref{thm:cone_s}, \ref{thm:cone_d}, and \ref{thm:cone_ic_s} 
can be directly extended to the nonlinear setting and provide a linear programming approach 
to nonlinear system analysis. This will be the objective of future publications. 

\appendix
\section{Proofs}\label{Appendix}

\subsection{Proof of Theorem~\ref{thm:cone_inv}}
Consider $x(0) \in \cX$ and $u(t) \in \cU$ for all $t \geq 0$. We show
that $x(t) \in \cX$ for all $t \geq 0$. 
By applying~\cite[Equation~(6)]{AS:03}, $x(t) \in \cX$ for all $t \geq 0$ if and only if
\begin{align*}
&x \in \cX, u \in \cU, \zeta \in \cX^*, \zeta^\top x=0\\
&\implies \quad
\zeta^\top ( A x + B u) \ge  0.
\end{align*}
Since the cone contains zero, the above condition can be decomposed into two conditions
\begin{align*}
x \in \cX, \zeta \in \cX^*, \zeta^\top x=0
\quad \implies \quad
\zeta^\top A x \ge  0.
\end{align*}
and
\begin{align*}
u \in \cU, \zeta \in \cX^*
\quad \implies \quad
\zeta^\top  B u \ge  0.
\end{align*}
By writing down these two conditions sing cone's generators, we get~\eqref{cond1:cone_inv} and~\eqref{cond2:cone_inv}, respectively.

Finally, we consider the condition for the outputs. which is equivalent to~$C \cX \subseteq \cY$ whenever $x(t) \in \cX$ for $t \geq 0$. 
Thus, \eqref{cond3:cone_inv} follows by writing down~$C \cX \subseteq \cY$ using cone's generators. 
\hfill \hspace*{1pt} \qed

\subsection{Proof of Theorem~\ref{thm:cone_s}}

(Sufficiency)
Let~$V(x)=v^\top x$. Condition~\eqref{cond1:cone_s} implies~$v \in \intr{\cX^*}$ and consequently~$V(x) > 0$ for all~$x \in \cX \setminus \{0\}$. According to~\eqref{cond2:cone_s}, i.e.,~$-v^\top A \in \intr{\cX^*}$, the time derivative of~$V(x)$ along the trajectory satisfies
\begin{align}
\frac{d V(x)}{dt} =  v^\top A x < 0
\end{align}
for all~$x \in \cX \setminus \{0\}$. Therefore, any trajectory starting from~$\cX$ converges to the origin. 
Furthermore, by linearity, both $x(\cdot)$ and $-x(\cdot)$ are system trajectories, therefore 
$-\cX$ is forward invariant and any trajectory starting from~$-\cX$ converges to the origin.
It remains to show that any other trajectory $z(\cdot)$ such that $z(0) \notin -\cX \cup \cX$ also converge to zero.
For instance, consider any trajectory $x(\cdot)$ such that $x(0)$ belongs to the interior of $\cX$ and define the trajectory 
$\eta(\cdot) = x(\cdot) + \varepsilon z(\cdot)$
where $0 < \varepsilon$ is small enough to guarantee $\eta(0) \in \cX$. Since $\eta(\cdot)$ converges to zero, also
$z(\cdot)$ must converge to zero. Exponential stability follows by linearity.

(Necessity)
Define
\begin{align*}
v 
&= \int_0^\delta e^{A^\top \tau} d { \tau}  \sum_{i_{\cX^*} \in \cI_{\cX^*}} X^*_{i_{\cX^*}}.
\end{align*}

From positivity ($e^{A \tau} X_{i_\cX} \in \cX\setminus \{0\}$,
for all $\tau \geq 0$, $i_{\cX} \in I_\cX$) and the fact that $\cX$ is a cone ($\cX + \cX \subseteq \cX$)
we have
\begin{align*}
\int_0^\delta e^{A \tau} d \tau X_{i_\cX} \in \cX \setminus \{0\}.
\end{align*}
This and~$\sum_{i_{\cX^*} \in \cI_{\cX^*}} X^*_{i_{\cX^*}} \in \intr{\cX^*}$ yield
\begin{align*}
v^\top X_{i_\cX} = \sum_{i_{\cX^*} \in \cI_{\cX^*}} (X^*_{i_{\cX^*}})^\top \int_0^\delta e^{A \tau} d \tau X_{i_\cX} > 0, \; \forall \delta \in \bR_+.
\end{align*}

The time derivative ~$v^\top \dot{x}$ at $x= X_{i_\cX}$,~$i_{\cX} \in \cI_\cX$ reads
\begin{align*}
v^\top A X_{i_\cX} = \sum_{i_{\cX^*} \in \cI_{\cX^*}} (X^*_{i_{\cX^*}})^\top (e^{A \delta} - I_n ) X_{i_\cX}.
\end{align*}
Since $\sum_{{i_{\cX^*}}} (X^*_{i_{\cX^*}})^\top X_{i_\cX} > 0$, for large~$\delta>0$ the right-hand side above is negative. 
\hfill \hspace*{1pt} \qed

\subsection{Proof of Theorems~\ref{thm:cone_d}}
(Sufficiency) 
Let~$V_s(x):=v^\top x \ge 0$ for all $x \in \cX$. 
From~\eqref{cond2:cone_d} and~\eqref{cond3:cone_d} we have
\begin{align*}
\frac{d V_s (x)}{dt} + \alpha V_s(x)
&= v^\top (A x + B u) + \alpha v^\top x \\
&\le q^\top y + r^\top u
\end{align*}
for all~$(x,u) \in \cX \times \cU$. This corresponds to \eqref{defn:ed}.

(Necessity)
Suppose that there exists a continuously differentiable function~$V_s(x) = v^\top x \ge 0$ for any $x\in \cX$, 
and~\eqref{defn:ed} holds,
that is,
\begin{align*}
v^\top(A x + B u) + \alpha v^\top x
& \leq q^\top C x + r^\top u
\end{align*}
for all~$(x,u) \in \cX \times \cU$. 
Define~$d: \cX \times \cU \to \bR_+$ as
\begin{align*}
d(x,u) = &-v^\top (A x \!+\! B u) - \alpha v^\top x
+  q^{\top} C x + r^{\top} u  \ge 0
\end{align*}
Take $d(x,\!u)= d_x(x) + d_u(x) u$ where
\begin{align*}
d_x(x) &= -v^\top A x - \alpha v^\top x +  q^\top C x, \\
 d_u(x) &=  -v^\top B + r^\top.
\end{align*}

Since $d(x,u) \ge 0$ for $(x,u) \in \cX \times \cU$ and $d_x(0) = 0$, 
it follows that~$d_u(0) u \ge 0$ for all~$u \in \cU$, i.e.,~$d_u(0) \in \cU^*$.
This gives \eqref{cond3:cone_d}.

Take now $u = 0$. Since $d(x,0) \ge 0$ for $x \in \cX $, we have $d_x(x) \geq 0$.
This gives,  \eqref{cond2:cone_d}.

Finally, \eqref{cond1:cone_d} follows directly from $V_s(x) \ge 0$ for any $x\in \cX$.
\hfill \hspace*{1pt} \qed

\subsection{Proof of Theorem~\ref{thm:cone_ic_s}}
First, we show $\cK$-positivity of the interconnected system,
\begin{align*}
\dot x_j = A_j x_j + B_j \sum_{k \in \cJ_N} W_{j,k} y_k, \; j \in \cJ_N.
\end{align*}
From a property of cone~$\cU_j + \cU_j \subset \cU_j$ and item~(I), it follows that~$\sum_{k \in \cJ_N} W_{j,k} \cY_k \subset \cU_j$. This and the cone invariance of the subsystem~$\Sigma_j$ implies~$x_j(\cdot ) \in \cX_j$,~$j \in \cJ_N$.

Next, we show exponential stability. Let~$V(x_1,\dots,x_N):= \sum_{j \in \cJ_N} \sigma_j v_j^\top x_j$, where~$V(x_1, \dots x_N) > 0$ for all $(x_1,\dots,x_N) \in (\cX_1 \times \cdots \times \cX_N) \setminus \{0\}$. From items~(II) and~(III), the time derivative of~$V(x_1,\dots,x_N)$ along the trajectory of the interconnected system can be computed as
\begin{align*}
&\frac{d V(x_1,\dots,x_N)}{dt}\\
&= \sum_{j \in \cJ_N} v_j^\top (A_j x_j + B_j u_j)\\
&\le - \sum_{j \in \cJ_N} \alpha_j v_j^\top x_j + \sum_{j \in \cJ_N} \sigma_j ( q_j^\top y_j + r_j^\top u_j)\\
&\le - \min_{j \in \cJ_N}\{ \alpha_j\}V(x_1,\dots,x_N)\\
&\hspace{5mm} + \sum_{j \in \cJ_N} \sigma_j  q_j^\top y_j +  \sum_{j \in \cJ_N}  \sigma_j  r_j^\top \sum_{k \in \cJ_N} W_{j,k} y_k \\
&\le - \min_{j \in \cJ_N}\{ \alpha_j\}V(x_1,\dots,x_N).
\end{align*}
Specifically, the first inequality above follows from (II); the last inequality follows from (III).

We can now  use Theorem~\ref{thm:cone_s} to establish the exponential stability of the interconnected system.
\hfill \hspace*{1pt} \qed

\bibliography{LC}                
\end{document}